# Understanding the different exciton-plasmon coupling regimes in two-dimensional semiconductors coupled with plasmonic lattices: a combined experimental and unified equations of motion approach

Wenjing Liu[1], Yuhui Wang[1], Carl H. Naylor[2], Bumsu Lee[1], Biyuan Zheng[3], Gerui Liu[1], A. T. Charlie Johnson[1,2], Anlian Pan[3], and Ritesh Agarwal[1*]

[1]Department of Materials Science and Engineering and [2]Department of Physics and Astronomy, University of Pennsylvania, Philadelphia, Pennsylvania 19104, United States

[3]Key Laboratory for Micro-Nano Physics and Technology of Hunan Province, School of Physics and Electronic Science, Hunan University, Changsha, Hunan 410082, P. R. China

* Email: riteshag@seas.upenn.edu

**ABSTRACT:** We study exciton-plasmon coupling in two-dimensional semiconductors coupled with Ag plasmonic lattices via angle-resolved reflectance spectroscopy and by solving the equations of motion (EOMs) in a coupled oscillator model accounting for all the resonances of the system. Five resonances are considered in the EOM model: semiconductor A and B excitons, localized surface plasmon resonances (LSPRs) of plasmonic nanostructures and the lattice diffraction modes of the plasmonic array. We investigated the exciton-plasmon coupling in different 2D semiconductors and plasmonic lattice geometries, including monolayer MoS$_2$ and WS$_2$ coupled with Ag nanodisk and bowtie arrays, and examined the dispersion and lineshape evolution in the coupled systems via the EOM model with different exciton-plasmon coupling parameters. The EOM approach provides a unified description of the exciton-plasmon interaction in the weak, intermediate and strong coupling cases with correctly explaining the dispersion and lineshapes of the complex system. This study provides a much deeper understanding of light-matter interactions in multilevel systems in general and will be useful to instruct the design of novel two-dimensional exciton-plasmonic devices for a variety of optoelectronic applications with precisely tailored responses.

Study of light-matter interactions is essential in understanding and manipulating the optical properties of materials to enable new and unprecedented functionalities. When light interacts with matter, in particular, a direct bandgap semiconductor, absorption of a photon leads to the formation of a coupled electron-hole pair (excitons) bonded via coulombic interaction. The exciton can then recombine radiatively and emit a photon, allowing energy transfer back and forth between the photon and the exciton. Depending on the relative magnitudes between such energy transfer rate (coupling strength, $g$), and the dissipation rate, $\gamma$ of each state, the system can be broadly classified into three light-matter coupling regimes[1]: (1) weak coupling regime, where the coupling strength $g \ll \frac{1}{2}(\gamma_e - \gamma_p)$, with $\gamma_e$ and $\gamma_p$ representing the decay rate of the excitonic and photonic states, respectively. In this regime, the eigenstates of the coupled system remain unchanged from their initial uncoupled states, and the system can be described by a perturbation theory where the Purcell effect[2], i.e., the modification of the spontaneous emission rate via engineering the photon density of states, can be observed. Purcell effect has been extensively studied for enhancing and suppressing the spontaneous emission rate in various cavity geometries[3-4], with applications in photonic and plasmonic lasers[5-6], brighter single-photon sources[7-8], hot luminescence[9-10] and quantum cryptography[11]. (2) intermediate coupling regime with $\frac{1}{2}(\gamma_e - \gamma_p) < g < \sqrt{\frac{1}{2}(\gamma_e^2 + \gamma_p^2)}$, in which normal mode splitting occurs in the frequency domain, and an anti-crossing behavior generally observed in the far-field optical spectrum. However, in time domain, the coherent energy transfer between the two states can be obscured by the large damping of one or more modes. (3) strong coupling regime, where $g > \sqrt{\frac{1}{2}(\gamma_e^2 + \gamma_p^2)}$, in which the energy can be coherently transferred between the two states for at least a few times before their eventual decay. Hence the system is characterized by new quasiparticles, called exciton-polaritons. As half-matter-half-light quasi-particles, polaritons possess the advantages of both the excitons and photons, such as enhanced inter-particle interactions, long coherent length and fast propagation speed, thus giving rise to numerous polaritonic devices including polariton lasers[12-14], switches[15], transistors[16], and optical circuits[17-18].

Light-matter interaction phenomena have been widely investigated in various geometries including both photonic and plasmonic cavities and numerous material systems.[19-20] Among different material systems, mono- and few-layered transition metal dichalcogenides (TMDs) have drawn great interest recently due to their unique optical[21-22] and electronic properties[23-25]. When thinned downed to a monolayer, they become direct band gap semiconductors with strongly bound excitons, owing to the reduced dielectric screening and strong quantum confinement in a 2D geometry[26-28], making them great testbeds to study light-matter interactions in 2D systems. Until now, numerous light-matter interaction phenomena have been demonstrated in these 2D semiconductors in different regimes, such as enhanced spontaneous emission[29-30], lasing[31], Fano resonance[29] and electromagnetically induced transparency (EIT)[32], and strong light-matter coupling[33-36], as well as active tuning between different light-matter coupling regimes[37-38], within various cavity geometries including diffractive Bragg reflectors[33], plasmonic nanocavities[39] and lattices[34, 36].

On the other hand, as the reduction of the optical mode volume is one of the critical routes to enhance the light-matter interaction strength, plasmonic systems, with their extraordinary ability to confine light far below the diffraction limit, providing useful platforms for enhancing light-matter interactions at the nanoscale[20, 40-42]. Especially, localized surface plasmon resonances (LSPRs), i.e., collective electron excitations at the surface of plasmonic nanostructures, can tightly confine and consequently strongly enhance the local electric field near the vicinity of these nanostructures, resulting in ultrasmall mode volumes and thus ultrastrong light-matter interactions.[43-45] When arranged into periodic arrays, LSPRs of individual nanostructures can couple effectively to the diffractive orders of the lattice near the Rayleigh-Wood's condition[46-47], i.e., the incident angle at which a diffractive beam passes off the array plane, giving rise to a unique type of plasmonic resonance called surface lattice resonance (SLR)[48-50]. Arising from coherent coupling between the LSPR of individual plasmonic nanoresonators and the lattice diffraction modes, SLR combines the advantages of both individual components, including strongly enhanced local E-field and small mode volume of the LSPRs, as well as enhanced quality factor, relatively long spatial coherence and high directionality of the propagating diffraction mode. Therefore,

SLRs serve as an excellent platform for studying and manipulating exciton-plasmon interactions, especially in 2D systems due to geometric compatibility.[51-54] Phenomenologically, SLRs can be explained by a coupled oscillator model (COM)[55], in which the LSPR and the lattice diffraction modes are treated as classical harmonic oscillators coupled via a phenomenological coupling strength, $g$. COM can reasonably predict both the mode dispersion and the lineshapes of the SLRs, providing beneficial understandings into the nature of SLRs and hence light-matter coupling in coupled plasmonic lattices.

Owing to the hybrid nature of SLRs, their optical properties, including mode volumes, resonance energies, dispersion and linewidths can all be engineered with great flexibility by tuning the LSPR and lattice diffraction modes independently via geometrical design, providing unique opportunities in tailoring exciton-plasmon coupling. In our previous work, by coupling monolayer $MoS_2$ with plasmonic lattices of different geometries, we have reported exciton-plasmon coupling in the weak, enhanced, and strong coupling regimes, featured by photoluminescence enhancement[29], Fano resonances[29], and normal mode splitting[34], respectively. However, these systems are quite complex due to the mutual coupling between multiple resonances, therefore precluding a straightforward and intuitive understanding of the exciton-plasmon coupling. In our recent studies of $MoS_2$-Ag nanodisk array system, for example, five resonance states are mutually coupled: A and B excitons, LSPR mode, and the lattice $\pm 1$ diffractive orders. In our study in the strong coupling regime, we applied a coupled oscillator model and solved the Hamiltonian matrix containing all the five coupled states to fit the measured angle-resolved dispersions, which provided valuable insights into the dispersion evolution while tuning the LSPR mode, as well as quantitative estimations of the light-matter coupling strengths in these complex systems.[34]

However, due to the limitation of the widely applied Hamiltonian matrix method, some critical questions still need to be addressed in the coupled MoS$_2$ exciton-plasmonic lattice system, and more generally, in any multilevel light-matter coupled system. Firstly, the Hamiltonian matrix calculates only the energies of the eigenstates yet provides little information about the lineshapes of the resonances. In plasmon-exciton systems, however, due to the interference between resonances with distinctly different linewidths, the resonance lineshapes can be significantly different from conventional Lorentzian lineshapes, that can manifest as asymmetric Fano resonances[56-57]. In addition to the resonance positions, characterization of the lineshapes along the dispersion are equally critical to understand the exciton-plasmon coupling in a coupled system. Previous studies on Fano resonances were mostly focused on two-level systems, while the resonance lineshapes in coupled multilevel systems have not yet been fully explored, where complex interference phenomena may be obtained. Secondly, in light-matter coupled systems with the coupling strengths comparable to the damping of the uncoupled states, or with large linewidth differences between different states, which typically occurs in plasmonic systems, the measured resonance positions in the frequency domain may not correspond exactly to the real eigenstate of the system, and the anti-crossing in dispersion may not truly imply strong coupling. In this case, the Hamiltonian matrix method may lead to inaccuracy in calculating the coupling strengths, and in extreme cases, may even result in wrong interpretations in determining the light-matter coupling regimes.[1, 58]

In order to evaluate these critical aspects, in this work, we have investigated the dispersion of exciton-plasmon coupling in 2D semiconductors coupled with plasmonic nanoresonator arrays by solving the equations of motion (EOM) involving the five different resonances of the system via a semiclassical coupled oscillator model.[59] The EOM method is a unified approach allowing seamless connection between all the different light-matter coupling regimes, which is desired in these highly tunable systems to fully understand the systems' properties and to guide geometrical designs for different applications. It allows the understanding of both the eigenenergies and the lineshapes of the coupled system, and has been successfully applied to various photonic and plasmonic systems to demonstrate spontaneous emission enhancement or suppression[1], Fano resonances [59-61], and strong light-matter coupling[62] but mostly in two-level systems with relatively simple dispersions. Furthermore, we will discuss the criteria of achieving the strong coupling limit in these multilevel systems, and conclude the discussion by extending the exciton-plasmon coupling concept to another monolayer semiconductor, WS$_2$, to show that our approach can be readily generalized to different 2D exciton-plasmon coupled systems. Besides, with larger oscillator strengths of the WS$_2$ excitons, the exciton-plasmon coupling is stronger, which enables us to push the system into the strong exciton-plasmon coupling regime, where $g > \sqrt{\frac{1}{2}(\gamma_e^2 + \gamma_p^2)}$.

In a semiclassical approach, the five resonances in the MoS$_2$-plasmonic lattice system (or in general, any exciton-plasmonic lattice system) can be treated as coupled harmonic oscillators with five coupled equations of motion written as [59]:

$$\ddot{x}_i + 2\gamma_i \dot{x}_i + \omega_i^2 x_i + 2\sum_j \varepsilon_{ij} g_{ij} \dot{x}_j = \delta_{LSPR} E(t) \qquad (1)$$

where the $i$ and $j$ =1-5 stand for the index of the five resonances; $x_i$ and $\gamma_i$ denote their amplitude and damping, respectively; $g_{ij}$ is the coupling strength between the $i^{th}$ and $j^{th}$ oscillators; $\varepsilon_{ij}$ is the Levi-Civita symbol which is defined as:

$$\varepsilon_{ij} = \begin{cases} +1 \text{ if } i < j \\ -1 \text{ if } i > j \\ 0 \text{ if } i = j \end{cases}$$

$E(t)$ is the driving force induced by the incident light with a given frequency $\omega$. Since the extinction cross section of the LSPR is much larger than the other resonances[59], we assume that only the LSPR is predominately excited by the external field, i.e., $\delta_{LSPR}$ becomes non-zero only if $i$ denotes the LSPR oscillator mode. Under a harmonic driving force $E(t) = E_0 e^{i\omega t}$, for steady state solutions, equation (1) can be simplified to:

$$(\omega_i^2 + 2i\gamma_i - \omega^2)x_i + 2i\omega \sum_j \varepsilon_{ij} g_{ij} x_j = \delta_{LSPR} E_0 \qquad (2)$$

We define $\mathbf{A_5}$ as the coefficient matrix of the five coupled oscillators obtained from equation (2), and $\mathbf{A_4}$ as the coefficient matrix of the four coupled oscillators except LSPR. The steady state solution of $x_{LSPR}$ can therefore be written as:

$$x_{LSPR} = \frac{Det(\mathbf{A_4}) E_0}{Det(\mathbf{A_5})} \qquad (3)$$

In our case, the damping of individual oscillators, detuning between any two states, and the coupling values are much smaller than the optical frequency, hence the system satisfies the near-resonance condition $\omega \sim \omega_i$, therefore, $\omega_i^2 - \omega^2 \approx 2\omega(\omega_i - \omega)$, and equation (3) can be further simplified to:

$$x_{LSPR} = \frac{\prod_{i=1}^{4}(\omega - \omega_i' - i\gamma_i')}{\prod_{i=1}^{5}(\omega - \omega_i - i\gamma_i)} \frac{E_0}{2\omega} \qquad (4)$$

where the denominator is a product of the Lorentzian factors of the complex eigenstates $(\omega_i + i\gamma_i)$ of all the five coupled oscillators calculated from the Hamiltonian matrix, while the nominator is composed of the eigenstates $(\omega_i' + i\gamma_i')$ of the other four coupled oscillators (except the LSPR mode). The extinction of the system can subsequently be calculated as the time-averaged work done by the incident light as $\langle E(t) \dot{x}_{LSPR}(t) \rangle \propto \omega \text{Im}(x_{LSPR})$.

In our experiments, silver plasmonic arrays with different nanoresonator sizes and shapes and lattice constants were patterned onto CVD grown $MoS_2$ (or $WS_2$) on 285 nm Si/SiO$_2$ substrate via e-beam lithography followed by physical vapor deposition (PVD) of 50 nm Ag (Figure 1 (a))[29, 34]. Reflectance spectra were measured by a home-built angle-resolved microscopy setup with the measured angle range defined by an objective of NA=0.7 and a white light source[63]. Figures 1 (b)-(d) present the typical experimentally measured reflectance spectra of bare $MoS_2$, plasmonic lattice (disk diameter, d = 135 nm; lattice constant, a = 460 nm), and the coupled $MoS_2$-SLR system (d =100 nm, a = 460 nm), and the data in (d) are adapted from ref 34. Since the LSPR wavelength red shifts by ~50-60 nm due to the local refractive index change when the plasmonic lattice is patterned onto $MoS_2$ (figures 1 (c) and (d)), the lattices were designed with different disk diameters to ensure that in both systems the LSPR is located at 640 nm, in resonance with the A exciton to allow direct comparison of the dispersion of the systems with the same LSPR wavelengths before and after coupling with $MoS_2$. In bare $MoS_2$, A and B exciton lines can be clearly observed at ~640 and ~590 nm, respectively with the damping $\gamma_A = 15$ meV and $\gamma_B = 30$ meV (figure 1 (b)), while in the plasmonic lattices, SLRs (yellow dashed lines) are formed due to the strong coupling between the LSPR (blue dashed line) and the lattice resonances (green dashed line), characterized by an anti-crossing centered at the crossing point of the two resonances (figure 1 (c)). When $MoS_2$ and the plasmonic lattice are coupled, as seen in figure 1 (d), the dispersion is strongly modified near the excitonic region, indicating significant exciton-plasmon interactions.

To investigate the exciton-plasmon interaction, we fitted the experimental data to the EOM described in equation (2). As we discussed in our previous work[34] and shown in figures 1 (b) and (c), the mode positions of the uncoupled states were identified as dips in the reflectance spectra, hence the system studied was absorption dominated, in which the dip (minimum) in reflectance corresponds to the maximum of extinction. Therefore, the experimental data were fitted to the negative extinction calculated by the EOM. It should be noted that scattering from plasmonic structures can influence the reflectance intensity, which will be discussed in more detail later. The EOM fitting results corresponding to figures 1 (b)-(d) are plotted in figures 1 (f)-(h), respectively.

For bare MoS$_2$ (figure 1 (f)), only the A and B excitons were considered in the EOM as two uncoupled oscillators, as the exciton-exciton interaction is assumed to be negligible in the system, and the driving force was applied to both excitons. In bare plasmonic lattice with the LSPR located at 640 nm (figure 1 (g)), three resonances were considered in the fitting: LSPR located at ~640 nm and the +1 and -1 lattice diffraction modes, with the fitted LSPR-lattice coupling strengths $g_{LSPR-lattice} = 100$ meV and negligible lattice-lattice coupling strength. The damping of the uncoupled LSPR and lattice diffraction modes could also be extracted by the EOM with $\gamma_{LSPR} = 100$ meV and $\gamma_{lattice} = 30$ meV. In the exciton-SLR coupled system (figure (h)), all five resonances were considered in the EOM, and the fitted light-matter coupling strengths are: $g_{A-LSPR} = 40$ meV, $g_{A-lattice} = 28$ meV, $g_{B-LSP} = 30$ meV, and $g_{B-lattice} = 20$ meV. It should be noted that a linear baseline was added to account for the finite band edge absorption of MoS$_2$ [64-65], which leads to a decrease of the total reflectance in the short wavelength region, as evident in the background reflectance spectrum in figure 1 (b). From figures 1 (f)-(h), it is worthwhile to notice that the EOM can nicely reproduce all the essential features in the reflectance spectra within large wavelength and angle ranges. Figure 1 (e) shows the EOM fitting results to the experimental data in figure 1 (d) in a series of line-cuts taken from different k$_{//}$ values of the angle-resolved reflectance spectrum, and a good agreement between the experimental and fitted data in both the resonance positions and lineshapes at all the measured k$_{//}$ values indicates that the EOM is a good model for this exciton-plasmon coupled system.

Although the overall fitting matches well with the experimental data, notable fitting errors appear close to $\sin\theta = 0$, where the reflectance intensity predicted by the EOM is lower than the experimental data. This deviation may result from two reasons: first, since the SLR arises from the coherent dipole-dipole interactions from individual plasmonic resonators, whereas the phenomenological coupled oscillator model treats the LSPR and the lattice diffraction mode as two coupled oscillators, and hence may not be able to fully reproduce the spectrum lineshape. As discussed in ref 55, when the LSPR is in close resonance with the lattice diffraction mode, the SLR lineshape can deviate from the Fano lineshape predicted by the EOM, resulting in only qualitative agreement. Such limitation of the model can also lead to errors in our present analysis. Second, in our analysis, the scattering contribution in the reflectance spectra is neglected, which may not always be negligible and hence gives rise to a higher reflection intensity. Indeed, a better lineshape fitting can be obtained by adding a scattering term $\sigma_{sca} \propto \omega^4 |x_{LSPR}|^2$ to the fitting, with the fitted dip positions and calculated coupling strengths remaining nearly unchanged. Therefore, to keep the model simple without significant sacrifice of the accuracy of the model, we will ignore scattering in the present work.

It is evident from figures 1 (d) and (h) that the exciton-plasmon coupling leads to complicated yet interesting lineshape modification in the reflectance dispersion measurements. Especially, although we have identified the dips as the resonance positions of the system, the reflectance spectrum in figure 1 (d) exhibits pronounced peak features (bright lines), which also display "anti-crossing" behavior near the A-exciton region, and these features are well reproduced by the EOM fitting. To understand this phenomenon and further the details of the exciton-plasmon coupling in the $MoS_2$-SLR system, we plotted a series of angle-resolved spectra calculated by the EOM while scanning one of the system's parameters to examine the evolution of the dispersion. Figures 2 (a)-

(e) present the extinction spectra as a function of the varying LSPR damping values $\gamma_{LSPR}$ = 5, 60, 100, 160, and 225 meV from (a) to (e), while the damping value in (c) corresponds to the experimental system in figures 1 (d) and (h). The rest of the parameters are kept as the same as in figure 1 (h), except that the wavelengths of the LSPR and A exciton are fixed exactly at 640 nm for simplicity, and the contribution from the B exciton was ignored. From figures 1 (a) to (e), it can be seen that as $\gamma_{LSPR}$ increases, the spectrum evolves from narrow and discrete dispersion lines in reflectance dips (dark lines) and undergoes linewidth broadening, and is eventually transformed to sharp bright peaks sandwiched between the broadband dark regions. To quantitatively characterize this phenomenon in detail, we first focus on the large angle region indicated by the blue dashed line in figure 2 (f) where $\sin\theta = 0.69$. At this angle, the two lattice diffraction modes are largely detuned from the A exciton, and hence the system can be treated as a two-level system composed of only the A exciton and LSPR. Such two-level coupled systems have been extensively studied and well understood in different light-matter coupling regimes. In this system, the oscillation amplitude of the LSPR in equation (4) is reduced to:

$$x_{LSPR} = \frac{\omega - \omega_A - i\gamma_A}{\prod_{i=1}^{2}(\omega - \omega_i - i\gamma_i)} \frac{E_0}{2\omega} \tag{5}$$

where $\omega_i + i\gamma_i$ stands for the two complex eigenstates of the A exciton-LSPR coupled system, and $\omega_A + i\gamma_A$ represents the complex resonance of the A exciton oscillator. We define $F = \left|\frac{1}{\omega - \omega_A - i\gamma_A}\right|^2$ and $S = \left|\frac{1}{\prod_{i=1}^{2}(\omega - \omega_i - i\gamma_i)}\right|^2$, which are the product of the Lorentzian resonances centered at the A exciton position and the eigenstates of the denominator of equation (5), respectively. Then equation (5) can be rewritten as:

$$|x_{LSPR}|^2 \propto \frac{S}{F} \tag{6}$$

We then plotted normalized S, F, and $|x_{LSPR}|^2$ associated with the parameters in figures 2 (a)-(e) in figures 2 (f)-(j). The *y*-axes of the plots were reversed to be consistent with the experimental data, hence the dips in the spectra correspond to the resonance positions. The vertical grey dotted lines represent the calculated eigenstates of the two-level system. For $\gamma_{LSPR} = 5$ meV (figure 2 (a)), the spectrum of S exhibits two distinct resonances located at the eigenstates of the strongly coupled t-level system, owing to the large coupling strengths relative to the damping. The curve F shows only a single dip corresponding to the A exciton resonance. Overall, the oscillator amplitude, $|x_{LSPR}|^2$, shows two resonances close to the system's eigenstates, with nearly symmetric line shapes, indicating that the system is in the strong coupling limit. In this limit, the overall spectrum is dominated by the S term, whereas the F term only alters the lineshapes of the resonances slightly. With the increase of $\gamma_{LSPR}$, the mode splitting between the eigenstates decreases and eventually disappears in figure 2 (h), and at the meantime, the system evolved progressively from strong to weak coupling limit. As a result, the resonances in S undergo linewidth broadening and merge into one resonance dip located at 640 nm, i.e., the wavelength of the uncoupled oscillators, whereas the spectrum of the F term remains unchanged. Consequently, as $\gamma_{LSPR}$ increases, the role played by the F term becomes significant: the relatively sharp dip in the F term leads to the suppression of the oscillation in the overall spectrum of $|x_{LSPR}|^2$ at the A exciton position, giving rise to a peak in the spectrum. As a result, the high- and low- energy dips in $|x_{LSPR}|^2$ gradually deviate from the corresponding eigenstates of the system, broaden and develop asymmetric lineshapes. This phenomenon is termed as Fano resonance, or electromagnetically induced transparency (EIT), resulting from the destructive interference between the sharp exciton resonance and the broadband LSPR mode. Therefore, as $\gamma_{LSPR}$ increases from 5 to 225 meV, the system evolves progressively from the strong coupling limit, characterized by normal mode splitting determined by the S term, to the intermediate/weak coupling limit, characterized by the Fano resonances determined by the F term.

We now examine the angle at which the lattice resonance is also at 640 nm so that the three oscillators: A exciton, LSPR, and one of the lattice diffraction modes are in resonance simultaneously (red dotted line in figure 2 (a)). Since the other lattice diffraction mode is largely detuned from 640 nm, the system can be approximated as a three-level system, where the oscillation amplitude of LSPR in equation 4 is reduced to:

$$x_{LSPR} = \frac{\prod_{i=1}^{2}(\omega-\omega_i'-i\gamma_i')}{\prod_{i=1}^{3}(\omega-\omega_i-i\gamma_i)} \frac{E_0}{2\omega} \qquad (7)$$

where $\omega_i + i\gamma_i$ and $\omega_i' + i\gamma_i'$ represent the complex eigenstates of this three-level system and the A exciton-lattice mode system, respectively. The F and S terms can now be written as $F = \left|\frac{1}{\prod_{i=1}^{2}(\omega-\omega_i'-i\gamma_i')}\right|^2$ and $S = \left|\frac{1}{\prod_{i=1}^{3}(\omega-\omega_i-i\gamma_i)}\right|^2$, which are the product of a group of Lorentzian resonances centered at the eigenstates corresponding to the numerator and denominator of equation (7), respectively. Therefore, S reflects the degree of coupling in the entire three-level system, while F represents the coupling between the A exciton and lattice mode.

The normalized S, F, and $|x_{LSPR}|^2$ corresponding to the red dotted line in figures 2 (a)-(e) are plotted in figures (k)-(o), respectively. Analogous to the A exciton-LSPR two-level system, with small LSPR damping $\gamma_{LSPR} = 5$ meV (figure 2 (a)), the $|x_{LSPR}|^2$ spectrum is dominated by the S term, displaying distinct, symmetric resonance dips near the three eigenstates of the system, while as $\gamma_{LSPR}$ increases, the system evolves from the strong coupling limit to the weak coupling limit, while the dip positions in the $|x_{LSPR}|^2$ spectrum progressively deviate from the system's eigenstates and the resonance lineshapes become asymmetric. Interestingly, when $\gamma_{LSPR} >$ 100 meV, S contains only one single resonance dip, and the $|x_{LSPR}|^2$ spectrum is mainly determined by the F term, which displays two distinct resonance dips that are independent of the value of $\gamma_{LSPR}$. These two dips give rise to two peaks in the $|x_{LSPR}|^2$ spectrum corresponding to oscillation suppression, which can be viewed as the Fano resonance in a mutually coupled three-level system. In other words, the angle-resolved spectrum in figure 2 (o) can be interpreted as follows: mathematically, it is possible to first allow the exciton-lattice coupling to form relatively sharp hybrid exciton-lattice states, which then interact with the broadband LSPR via Fano interference, leading to the formation of the sharp peaks in the spectrum which exhibit "anti-crossing" behavior in their dispersion induced by the initial exciton-lattice coupling. Between the strong coupling limit dominated by the S term (figure 2 (a)) and the weak coupling limit dominated by the F term (figure 2 (e)), the experimental system (figure 2 (d)) is somewhere in the middle of the two limiting cases and is influenced by both mechanisms: resonances (dips) near the system's eigenstates were observed in the $|x_{LSPR}|^2$ spectrum caused by normal mode splitting, but with slight deviations in the resonance positions and asymmetry in the lineshapes resulting from Fano interference.

As discussed in figures 1 and 2, although the resonances of the MoS$_2$-plasmonic lattice system present as dips in the reflectance spectrum, the "peaks" in the spectrum also display "anti-crossing" behavior near the excitons in their dispersions, which relate closely to the exciton-lattice diffraction mode coupling, although these peaks do not correspond directly to the actual resonance positions of the system. This interesting phenomenon is further confirmed in figures 3 ((a) – (l)), in which $g_{A-lattice}$ was scanned from 0 to 60 meV, with $g_{A-lattice} = 28$ meV in figure 3 (c) corresponding to the fitted experimental value, while all the other parameters are fixed to the realistic values in figures 1 (d) and (h) and figure 3 (c). Indeed, the "mode splitting" value between the peaks increases significantly as $g_{A-lattice}$ increases. It can also be resolved from the line-cut plots in figures 3 (f)-(j) that the peaks in $|x_{LSPR}|^2$ match well with the resonances in the spectrum of F, which is an indication of the degree of the A exciton-lattice mode coupling. Moreover, the role played by $g_{A-LSP}$ in modifying the system's dispersion was also investigated in figure 4 by scanning $g_{A-LSP}$ from 0 to 100 meV (Figures 4 (a)-(e)) while fixing other parameters. While $g_{A-lattice}$ determines the spectral evolution near the crossing positions between the lattice and the A exciton, $g_{A-LSP}$ dominates near high $k_{//}$ and $k_{//}$ values close to 0, where the lattice mode is largely detuned from the A exciton. Analogous to figures 2 (f)-(j), at $\sin\theta = 0.7$, the lattice modes are detuned by more than 300 meV from the A exciton, hence the system can be approximately treated as a two-level system including only the A exciton and the LSPR, and the evolutions of S, F, and $|x_{LSPR}|^2$ with $g_{A-LSP}$ in this two-level system are plotted in figures 4 (f)-(j). When $g_{A-LSP} = 0$ meV, the A exciton is fully decoupled from the LSPR mode (figures 4 (a) and (f)), hence the spectrum presents only one broad dip at 640 nm. As $g_{A-LSPR}$ increases, a sharp peak shows up at 640 nm due to the Fano interference (figures 4 (b) and (g)), implying enhanced exciton-plasmon coupling. Upon further increasing $g_{A-LSP}$, in figures 4 (c) and (h), when $g_{A-LSP} > \frac{1}{2}(\gamma_{LSPR} - \gamma_A)$, the eigenstates of the system start to split, indicating that the system enters the intermediate coupling regime. The system reaches the strong coupling regime in figures 4 (d) and (i), where $g_{A-LSPR} > \sqrt{\frac{1}{2}(\gamma_{LSPR}^2 + \gamma_A^2)}$, which is also evident by the splitting in the spectrum of S. Further increase of $g_{A-LSP}$ results in the increase of the mode splitting in both S and $|x_{LSPR}|^2$ (figures 4 (e) and (j)).

In the above discussion, we investigated the evolution of exciton-plasmon coupling dispersion and lineshapes in the MoS$_2$-plasmonic lattice system with different parameters in a EOM model, where the LSPR is in resonance with the A exciton. In this system, due to the complex SLR dispersion, exciton-plasmon coupling can have distinctive characteristics controlled by different light-matter coupling parameters at different angles. Moreover, as a result of the mutual coupling between oscillators with different linewidths, Fano lineshapes have been extensively observed in weak, intermediate and even strong coupling regimes, which leads to an interesting consequence that "anti-crossing" behaviors can be observed in both dip (resonance) and peak dispersions. As we have learnt that the LSPR plays a critical role in manipulating the exciton-plasmon coupling dispersions and lineshapes, therefore, to further investigate the influence of the LSPR as it is tuned in and out of resonance with the excitons, four sets of Ag nanodisk arrays were patterned onto the $MoS_2$ monolayers, with the same lattice pitch of 460 nm and different nanodisk diameters varying from 70 nm – 150 nm to allow the LSPR resonance to span across the excitonic region of $MoS_2$. The measured angle-resolved dispersions and their corresponding EOM fittings are presented in figure 5 (data was adapted from ref 34). The EOM was capable to reproduce the experimental data for all four arrays over a wide wavelength and angle range, and the exciton-plasmon coupling strength values obtained from EOM fitting are listed in Table 1. Comparing with our earlier work in which the coupling strength values were obtained from diagonalizing the Hamiltonian matrix in a coupled oscillator model, the evolution of the extracted coupling strengths calculated by the EOM follows qualitatively the same trend, while the absolute values can differ by ~10-30%, depending on the LSPR-A exciton detuning. As discussed above (figures 2, 3 and 4), this difference mostly arises because that the experimentally measured resonance energies in the optical spectrum does not necessarily correspond exactly to the energies of the intrinsic eigenstates of the coupled system, therefore the EOM approach gives more accurate estimations of the light-matter coupling strengths than the Hamiltonian matrix method. On the other hand, the dispersion lineshapes also evolve progressively with the LSPR-exciton detuning. In lattices with disk diameters of d=70 nm and d=100 nm (figures 5 (a) and (c)), the LSPR is in close resonance with the B or A excitons, as a result, the angle-resolved spectra display sharp peak-like dispersions with anti-crossing behaviors as discussed in figure 2. On the contrary, as the LSPR is gradually detuned from the excitonic region towards long wavelengths (figures 5 (e) and (g)), the sharp peak features resulting from Fano resonances gradually broaden and diminish, and in figure 5 (g), where the LSPR-A exciton detuning is more than 300 meV, the system recovers typical anti-crossing behavior in the dip dispersions resulting mostly from the exciton-lattice diffraction mode coupling.

**Table 1** Coupling strengths between the five resonances as a function of Ag nanodisk diameter in monolayer MoS$_2$ coupled with Ag nanodisk array.

| Ag nanodisk diameter | $g_{A-LSPR}$ (meV) | $g_{A-Lattice}$ (meV) | $g_{B-LSPR}$ (meV) | $g_{B-Lattice}$ (meV) | $g_{LSPR-Lattice}$ (meV) |
|---|---|---|---|---|---|
| d=70 nm | 34 | 20 | 31 | 15 | 70 |
| d=100 nm | 39 | 28 | 28 | 17 | 87 |
| d=120 nm | 35 | 27 | 25 | 18 | 110 |
| d=150 nm | 25 | 25 | 20 | 15 | 128 |

Now we will discuss the nature of the exciton-SLR coupling in the MoS$_2$-plasmonic lattice structures presented in figure 5, i.e., the light-matter coupling regimes of these systems. In different light-matter coupling regimes, the spontaneous emission rate, energy transfer, and inter-particle interactions can be significantly altered, leading to very distinct optical phenomena. Therefore, the identification of the coupling regime is essential in understanding the nature of the light-matter interaction that can help guide the design of practical devices. Due to the complexity of the mutually coupled multi-level system, a simple classification of the exciton-plasmon coupling may not be readily applied. However, under certain conditions, the system may be approximately simplified to a two-level system with only one exciton and one SLR mode, where the nature of the light-matter coupling can be examined. We will first focus on the spectrum in figures 5 (e) and (f) as an example. The SLR dispersion of the system can be identified by removing the coupling between the excitons and the plasmonic resonances ($g_{A-lattice}$, $g_{B-lattice}$, $g_{A-LSP}$ and $g_{B-LSP}$) from the EOM, as plotted in figure 6 (a), where only the plasmonic modes are presented. This dispersion cannot be obtained directly from experimental measurements because of the red-shift of the LSPR when the Ag nanodisk is patterned onto $MoS_2$ compared with nanodisks patterned on the $SiO_2$/Si substrate. From this calculated dispersion, we can locate the angle at which the SLR mode is in resonance with the A exciton, as indicated by the crossing point of the yellow and the white curves in figure 6 (a). At this angle ($\sin\theta = 0.45$), the lower SLR branch is detuned from the exciton by more than 300 meV, and B exciton is detuned from A by ~160 meV, hence their influences can be ignored, and the system can be approximately treated as a two-level system. By fitting the dispersion within the region enclosed by the red dashed box to the EOM of two coupled oscillators (figures 6 (b) and (c)), we obtain the A exciton-SLR coupling strength $g_{A-SLR} = 34$ meV, and the damping of the SLR $\gamma_{SLR} = 61$ meV, which is greatly reduced from the original (uncoupled LSPR) $\gamma_{LSPR} = 100$ meV due to its coupling with the lattice mode.

The criterion of having normal mode splitting in two-level systems in frequency domain at zero-detuning conditions can be written as:

$$g > \frac{|\gamma_1 - \gamma_2|}{2} \tag{8}$$

Which is fulfilled in the system investigated in figure 5 (c), and all other three systems presented in figures 5 (a), (e) and (g) as well, indicating the occurrence of eigenstate splitting resulting from exciton-SLR coupling in these plasmonic lattices. However, the eigenstate splitting in frequency domain does not necessarily imply strong coupling in time domain, i.e., the observation of a few Rabi oscillations between the excitonic and photonic modes before the system decays. In order to reach the strong coupling regime, a stricter condition needs to be satisfied so that at least one Rabi cycle can be completed before the decay of the excitations, i.e.,

$$g^2 > \frac{\gamma_1^2 + \gamma_2^2}{2} \tag{9}$$

In the system discussed in figure 2 (c), $g_{A-LSP}^2 < \sim \frac{\gamma_A^2 + \gamma_{SLR}^2}{2}$, implying that we are approaching the strong coupling limit. The criterion applied here is the strictest to determine if the system is in the strong coupling regime, and hence other criteria are automatically fulfilled.

We will now revisit the exciton-plasmon coupling in bowtie arrays that we investigated previously[29]. In this work, we observed Fano resonances near the exciton positions with certain bowtie geometries. Here, with a more comprehensive understanding of the light-matter interactions in the $MoS_2$-plasmonic lattice systems, by applying the EOM approach, we present a quantitative description of the exciton-plasmon coupling in these $MoS_2$-bowtie array systems. A set of Ag bowtie arrays were fabricated onto the $MoS_2$ monolayers with the same bowtie sizes (triangle side, s=100 nm; gap, g = 20 nm; thickness, h=50 nm) and different lattice constants ranging from 350 – 550 nm. The measured angle-resolved reflectance spectra of the different arrays and their corresponding EOM fitting are presented in figures 7 (a), (c), (e) and (b), (d), (f), respectively. The reflectance spectra measured from the MoS$_2$-bowtie structures were reasonably reproduced by the EOM for all the three samples, with the LSPR of the bowtie nanoresonator identified at ~645 nm. The exciton-plasmon coupling strengths in these bowtie arrays are listed in Table 2. Comparing with Ag nanodisk devices, the LSPR of bowties exhibits a much larger damping (~150 meV), while giving rise to slightly larger coupling strengths with excitons, for mainly two reasons: (1) comparing with the dipolar mode of the nanodisk, bowtie supports a gap plasmonic mode that is tightly confined between its two tips, which results in a larger field confinement at the expense of a higher damping; (2) the LSPR of bowties are more sensitive to fabrication errors, which can lead to a larger inhomogeneous broadening of the LSPR mode. Hence in our design, nanodisks are better nanostructures for pursuing strong exciton-plasmon coupling. The exciton-LSPR coupling strength in the MoS$_2$-Ag bowtie device increases as the lattice constant decreases, probably due to the larger density of the plasmonic nanoresonators in arrays with smaller pitch. Especially, in figures 5 (a) and (b), the geometrical parameters of the array are similar to that in figure 5 (a) of ref 29, where the most pronounced Fano resonance was observed. In this geometry, the lattice mode is largely detuned from the excitonic region, and the system is dominated by only the exciton-LSPR coupling. In the measured spectrum, a peak is observed close to the A exciton energy due to electromagnetically induced transparency, and can be clearly resolved in the line-cut plot (figure 5 (g)) taken at $\sin\theta = 0$ from figure 5 (a), which is consistent with our previous observations in ref 29. Comparing the exciton-LSPR coupling strengths with the damping of the exciton and the LSPR, we obtain $g_{A-LS} \sim < \frac{1}{2}(\gamma_A - \gamma_{LSPR})$, indicating that the system is approaching the intermediate coupling regime, hence the transparency is induced mainly by the Fano interference between the exciton and the LSPR.

**Table 2**. A-exciton-plasmon coupling strengths in Ag bowtie arrays with different lattice constants.

| plasmonic lattice constant | $g_{A-LSPR}$ (meV) | $g_{A-Lattice}$ (meV) | $g_{B-LSPR}$ (meV) | $g_{B-Lattice}$ (meV) | $g_{LSPR-Lattice}$ (meV) |
|---|---|---|---|---|---|
| a=350 nm | 39 | 20 | 30 | 18 | 110 |
| a=450 nm | 37 | 21 | 28 | 15 | 100 |
| a=550 nm | 30 | 20 | 25 | 14 | 70 |

After assigning the exciton-plasmon coupling in MoS$_2$-plasmonic arrays, we now generalize the concept to other monolayer semiconductors. We chose monolayer WS$_2$ as the active material because it exhibits very strong A exciton features near 620 nm with large oscillator strengths.[27] We patterned Ag nanodisk arrays with disk diameters of 100 nm and lattice constant of 430 nm on monolayer WS$_2$, and investigated the temperature dependent WS$_2$ exciton-SLR coupling (Figure 8). The WS$_2$ sample in our study was grown by CVD on 300 nm SiO$_2$/Si substrate, exhibiting strong and narrow A exciton resonances even up to room temperature (figure 8 (a)). We performed temperature dependent angle-resolved reflectance measurement on the WS$_2$-plasmonic lattice samples at 77, 200 and 300K, and the results are shown in figures 8 (b)-(d) along with the corresponding EOM fitting results in figures 8 (e)-(g). As a result of the large exciton oscillator strength in WS$_2$, a significant exciton-plasmon coupling was observed up to room temperature, with only a slight reduction in $g_{A-LSP}$ values from 44 meV at 77 K to 34 meV at 300 K. It is worth noting that at 77 K, the exciton-plasmon coupling as calculated from the aforementioned method exceeds the strong coupling limit, $g_{A-LS}^2 > \frac{\gamma_A^2 + \gamma_{SLR}^2}{2}$ indicating that indeed strong exciton-plasmon coupling was realized in the WS$_2$-SLR system. This finding hence holds promises for designing exciton-plasmon polariton devices in 2D semiconductors coupled with plasmonic lattices.

To conclude, we have studied light-matter interactions in monolayer TMD semiconductors coupled to plasmonic lattices in various lattice geometries. By solving the equations of motion in a coupled oscillator model accounting for the five different resonances of the system, we provided a unified description of the exciton-plasmon coupling in the weak, intermediate and strong coupling regimes in these complex, multi-level systems. The lineshape evolution of the system as the function of the light-matter coupling parameters and different exciton-plasmon coupling regimes were also discussed to provide insights about the system. By utilizing monolayer WS2, a 2D semiconductor with excitons with large oscillator strengths, we have demonstrated strong light-matter coupling and formation of polaritons in the system. The generalized description of exciton-photon coupling in a multi-level system can be utilized to understand and engineer light-matter interactions in other excitonic systems with precisely engineered response for a variety of optoelectronic applications

**Acknowledgements**: This work was supported by the NSF under the NSF 2-DARE program (EFMA-1542879), NSF-MRSEC (LRSM) seed grant under award number DMR11-20901 and by the US Army Research Office under Grant No. W911NF-12-R-0012-03. A.P. is grateful to the National Natural Science Foundation of China (No.51525202), the Aid Program for Science and Technology Innovative Research Team in Higher Educational Institutions of Hunan Province. Nanofabrication and electron microscopy characterization was carried out at the Singh Center for Nanotechnology at the University of Pennsylvania.

**FIGURE CAPTIONS**

**Figure 1. Optical properties of monolayer MoS₂, plasmonic nanodisk lattice, and coupled MoS₂-plasmonic lattice system and the corresponding EOM fitting.** (a) SEM image of a typical silver nanodisk lattice patterned on monolayer MoS₂. (b) Angle-resolved reflectance spectrum of monolayer MoS₂ at 77 K. A and B excitons are indicated by white dashed lines. (c) Angle-resolved reflectance spectrum of Ag nanodisk array with disk diameter of 135 nm and a lattice constant of 460 nm. green dashed line denotes the lattice diffraction mode, blue dashed line represents the LSPR of the silver nanodisk, and yellow dashed line corresponds to the coupled SLRs of the lattice. (d) Angle-resolved reflectance spectrum of monolayer MoS₂ coupled with silver nanodisk array with disk diameter of 100 nm and a lattice constant of 460 nm. (e) The line-cut spectra taken from (d) $\theta = -4°$ to $\theta = -25.9°$ from bottom to top and their corresponded EOM fitting. (f)-(h) EOM fitting results of (b), (c) and (d), respectively. (d) data adapted with permission from ref. 34. Copyright © 2016 American Chemical Society.

**Figure 2. Evolution of the dispersion of the MoS₂-plasmonic lattice system with increasing $\gamma_{LSPR}$ obtained by the EOM model.** (a)-(e) Angle-resolved extinction spectrum calculated by the EOM model with $\gamma_{LSPR} = 5, 60, 100$ and $160 \text{ meV}$ and $250 \text{ meV}$, respectively. Other parameters fixed at: $g_{A-LSPR} = 40 \text{ meV}$, $g_{A-lattic} = 28 \text{ meV}$, and $g_{LSPR-lattic} = 100 \text{ meV}$ (c) Simulation parameters corresponding to the experimentally measured spectrum in figure 1 (d). (f)-(j) Normalized S, F and $|x_{LSPR}|^2$ corresponding to (a)-(e), respectively,

taken from the angle indicated by the blue dashed line in (a). (k)-(o) Normalized S, F and $|x_{LSPR}|^2$ corresponding to (a)-(e), respectively, taken from the angle indicated by the red dashed line in (a).

Figure 3. Evolution of the dispersion of the MoS₂-plasmonic lattice system with increasing $g_{A-lattice}$ (A exciton-lattice coupling) obtained by the EOM model. (a)-(e) Angle-resolved extinction spectra calculated by the EOM model with $g_{A-lattice} = 0, 15, 28,$ and 45 meV, and 60 meV, respectively. Other parameters are fixed at: $\gamma_{LSPR} = 100$ meV, $g_{A-LSP} = 40$ meV, and $g_{LSPR-lattice} = 100$ meV (c) simulation parameters corresponding to the experimentally measured spectrum in figure 1 (d). (f)-(j) Normalized S, F and $|x_{LSPR}|^2$ corresponding to (a)-(e), respectively, taken from the angle indicated by the red dashed line in (a).

Figure 4. Evolution of the dispersion of the MoS₂-plasmonic lattice system with $g_{A-LSPR}$ (A exciton-LSPR coupling) obtained by the EOM model. (a)-(e) Angle-resolved extinction spectra calculated by the EOM model with $g_{A-LS} = 0, 30, 40, 70$ and 100 meV, respectively. Other parameters are fixed at: $\gamma_{LSPR} = 100$ meV, $g_{A-lattice} = 28$ meV, and $g_{LSPR-lattice} = 100$ meV, and (c) simulation parameters corresponding to the experimentally measured spectrum in figure 1 (d). (f)-(j) Normalized S, F and $|x_{LSPR}|^2$ corresponding to (a)-(e), respectively, taken from the angle indicated by the red dashed line in (a).

**Figure 5. Evolution of the dispersion of the MoS$_2$-nanodisk array system with varying Ag nanodisk diameters.** (a), (c), (e), (g) Experimentally measured angle-resolved spectra of MoS$_2$ coupled with nanodisk arrays with disk diameters of (a) 70 nm, (c) 100 nm, (e) 120 nm, (g) 150 nm. Adapted with permission from ref. 34. Copyright © 2016 American Chemical Society. Corresponding EOM fitting are shown in (b), (d), (f) and (h) respectively.

**Figure 6. A exciton-SLR coupling near zero A exciton-SLR detuning region of the dispersion.** (a) SLR dispersion obtained by the EOM fitting corresponding to the spectrum in figures 5 (e) and (f). Green and blue dashed lines represent the uncoupled lattice diffraction modes and LSPR, yellow dotted lines represent the SLR dispersion calculated by EOM fitting, and white dashed line represents the A exciton. (b) and (c) Experimentally measured angle-resolved dispersion close to zero A exciton-SLR detuning point, with the wavelength and angle range indicated by a dashed red box in (a). (c) Two-level EOM fitting of (b). (d) Line-cut spectra taken from (b) and (c).

**Figure 7. Evolution of the dispersion of the MoS$_2$-Ag bowtie array system with varying lattice constant of the Ag nanodisks.** (a), (c) and (e) Experimentally measured angle-resolved spectra of MoS$_2$ coupled with Ag bowtie arrays with lattice constants of (a) 350 nm, (c) 450 nm and (e) 550 nm. (b), (d) and (f) EOM fitting corresponding to (a), (c) and (e), respectively. (g) line-cut spectrum taken from $\theta = 0$ in (a) and the bare MoS$_2$ reflectance spectrum

measured from the same MoS$_2$ flake.

**Figure 8. Evolution of the optical properties of the WS$_2$-nanodisk array system with temperature.** (a) Reflectance spectrum measured from bare monolayer WS$_2$ at 77, 200 and 300 K. (b), (d) and (f) Experimentally measured angle-resolved spectra of WS$_2$ coupled with Ag nanodisk arrays with temperatures of (b) 77 K, (d) 200 K and (f) 300 K. (c), (e) and (g) EOM fitting results corresponding to (b), (d) and (f), respectively.

Figure 1.

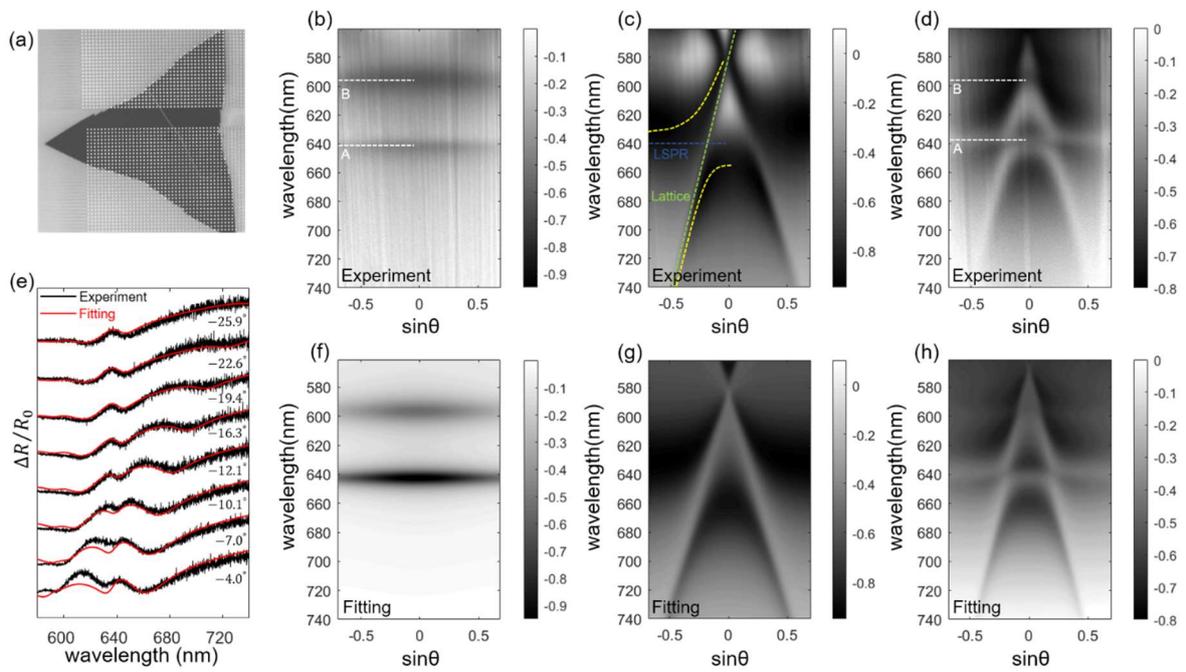

Figure 2.

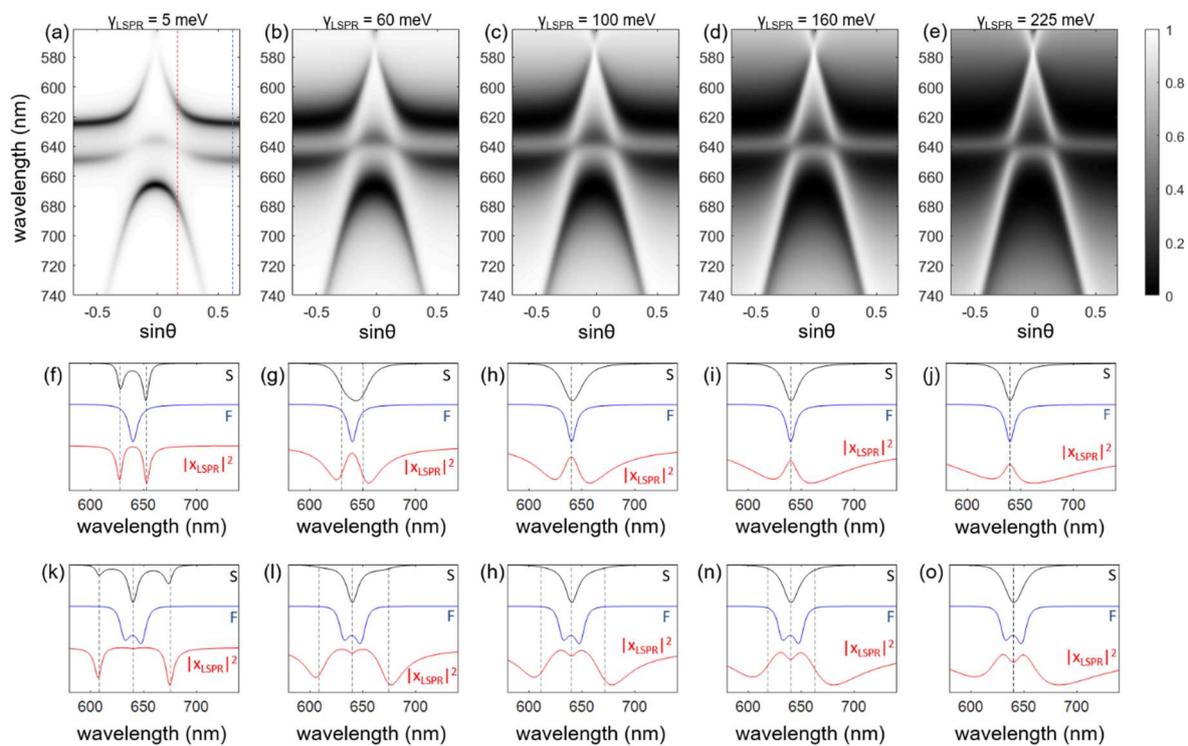

Figure 3.

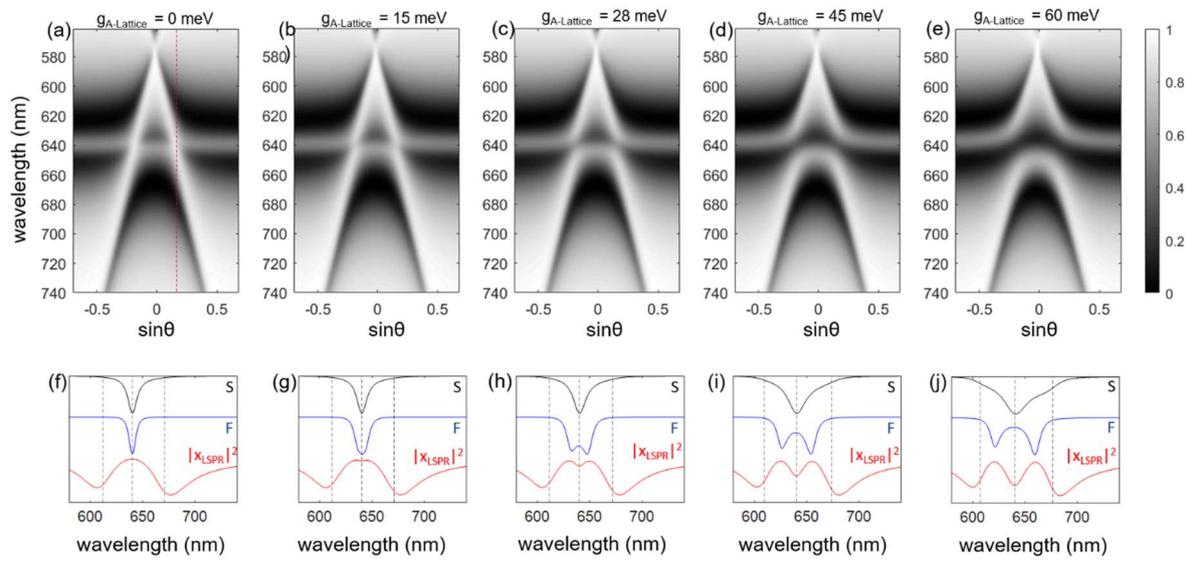

Figure 4.

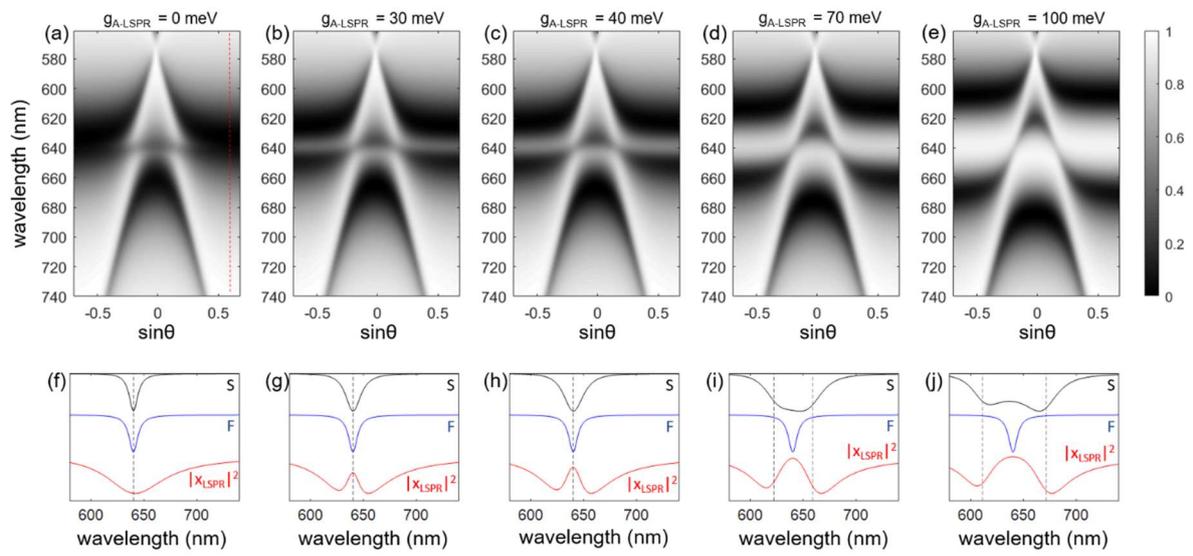

Figure 5.

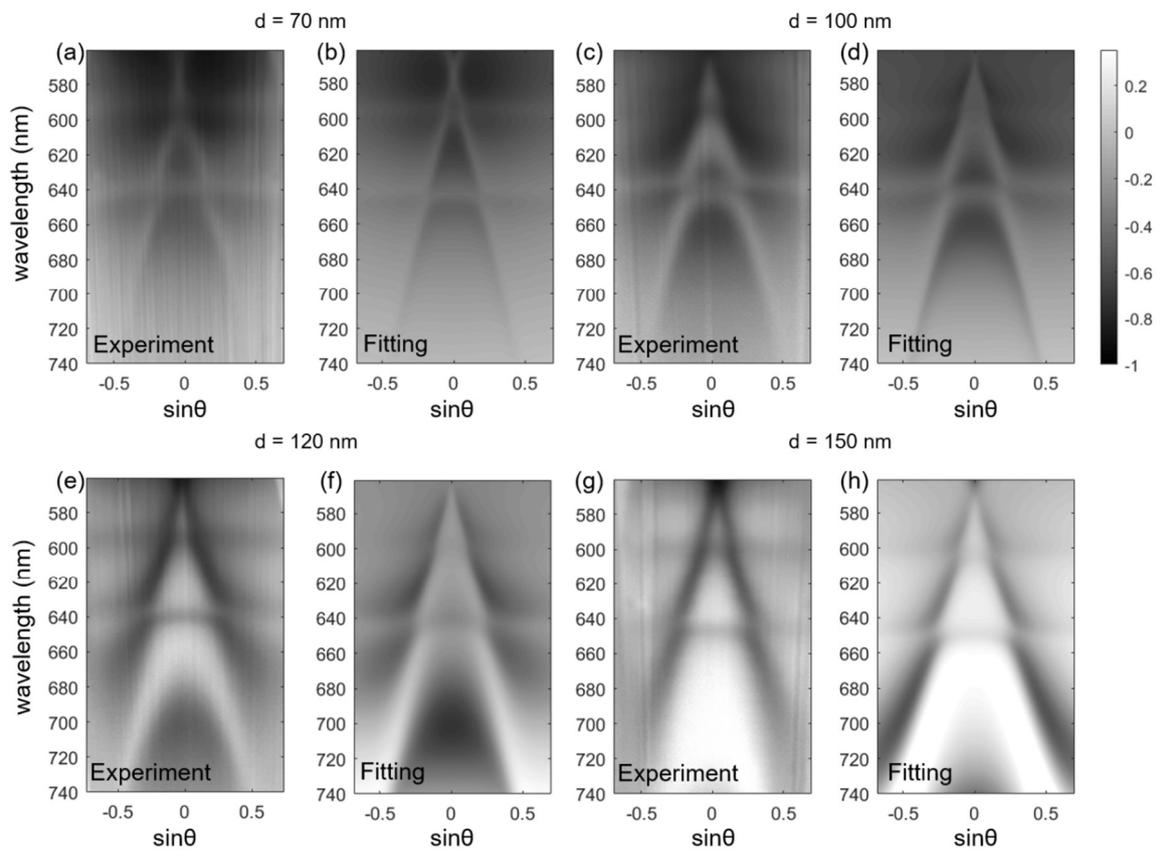

Figure 6.

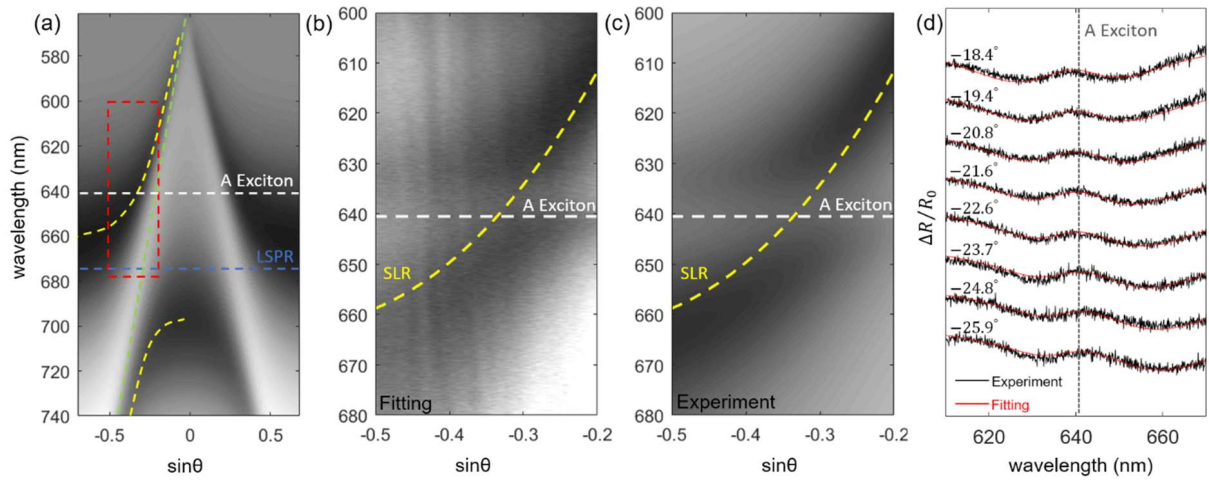

Figure 7.

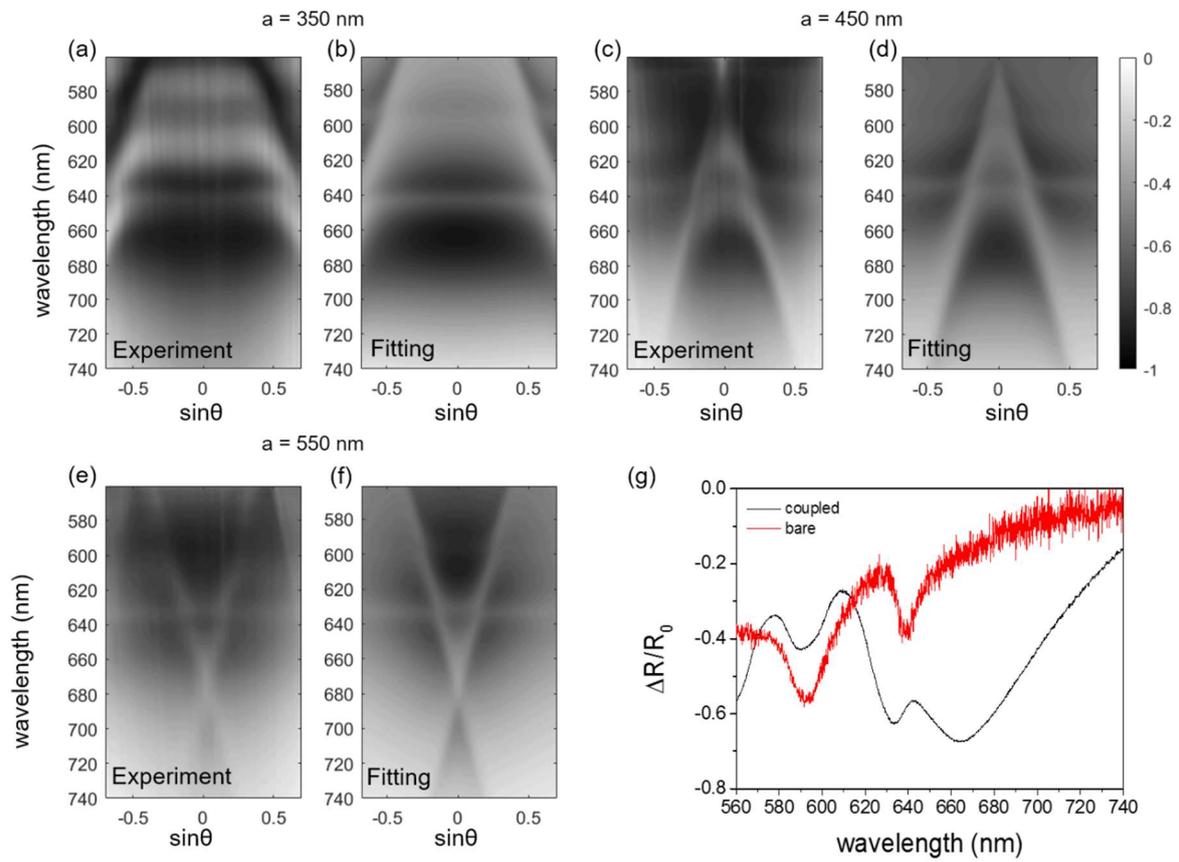

Figure 8.

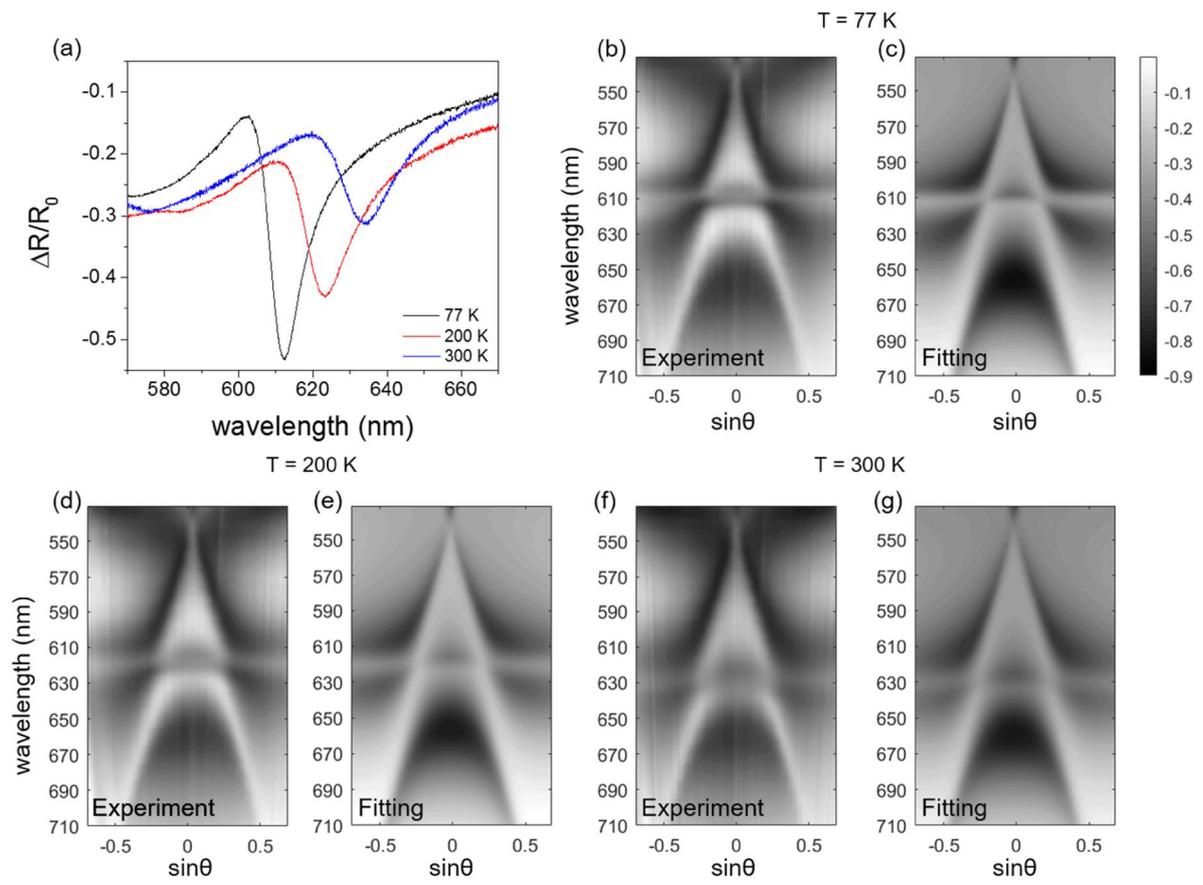

TOC:

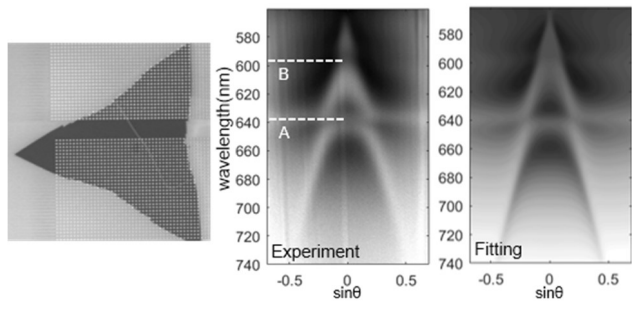